\documentclass[12pt,psf,epsf]{article}
\usepackage[centertags]{amsmath}
\usepackage{amssymb}
\usepackage{graphicx}
\usepackage{epsfig}
\usepackage{ulem}
\usepackage[english]{babel}
\usepackage{array}
\usepackage{amsthm}
\usepackage{latexsym}
\usepackage[mathcal]{euscript}
\pdfoutput=1
\usepackage{epsfig}
 \usepackage{jheppubm}
\usepackage{mathdots}
\usepackage{MnSymbol}
\usepackage{multirow}

\newcommand{\nn}{\nonumber}
\newcommand{\be}{\begin{equation}}
\newcommand{\ee}{\end{equation}}
\newcommand{\bea}{\begin{eqnarray}}
\newcommand{\eea}{\end{eqnarray}}

\newcommand{\e}{\mathrm{e}}

\newcommand{\Tr}{\text{Tr}}

\definecolor{dgreen}{rgb}{0.,0.6,0.}

\title{$$\,$$\\
$$\,$$\\Spontaneous symmetry breaking in fermionic random matrix model}
\author{Irina Aref'eva and}
\author{Igor Volovich}

\affiliation{Steklov Mathematical Institute, Russian Academy of Sciences,\\Gubkina str. 8, 119991, Moscow, Russia}

\abstract{
A fermionic random matrix model,   which is a  0-dimensional version of the  SYK model with  replicas,   is considered.  
The replica-off-diagonal correlation functions vanish at finite $N$, but we show that they do not vanish in the large $N$ limit due to spontaneous symmetry breaking. We use  the Bogoliubov  quasi-averages approach to studying phase transitions.  The consideration may be relevant to the study of the problem of existence of the spin glass phase in fermionic models.
}

\emailAdd{arefeva@mi-ras.ru}

\emailAdd{volovich@mi-ras.ru}
\begin{document}
\maketitle

\newpage
\section{Introduction}

Spontaneous symmetry breaking is a common occurrence in solid state and high energy physics including the Ising model, Bose-Einstein condensation, Higgs mechanism and  spin glasses \cite{Landau,Bog, Numbu,Weinb,MezPar}. In these cases the stable solutions of the equations, which govern the system, exhibit less symmetry than the equations themselves. One has a quantum state that is invariant under a symmetry group and which admits a nontrivial decomposition into extremal states  corresponding to pure thermodynamic phases. Usually spontaneous symmetry breaking shows in the range of parameters where the phase transitions take place.

In this note a fermionic random matrix model \cite{9auth,SSS,AreVol} is considered which is a  0-dimensional version of the  SYK model \cite{SY,AK,Sachdev:2015} with a finite  number of replicas. 
The replica-off-diagonal correlations functions vanish at finite $N$, but we show that they do not vanish in the large $N$ limit due to spontaneous symmetry breaking. We use  the Bogoliubov  quasi-averages approach \cite{Bog,Zagr} to studying phase transitions. 

We consider a model of fermions $\psi_j^{\alpha}$ where $j=1,...,N,\,\alpha =1,...,M$ with the Hamiltonian or action which are invariant under the "gauge group" transformations $\psi_j^{\alpha}\to\epsilon^{\alpha}\psi_j^{\alpha}$ where $\epsilon^{\alpha}=\pm 1$. It follows that for finite $N$ the 2-point off-diagonal correlation function vanishes $\langle \psi_j^{\alpha}\psi_j^{\beta}\rangle_N =0,\,\alpha\neq\beta$. One can also argue that the  2-point correlation function for finite $N$ vanishes, since it  factorizes into fermion 1-point functions which are zero. However, we will show that in the fermionic random matrix model in the limit $N\to\infty$ there is spontaneous symmetry breaking  and the 2-point off-diagonal correlation function is non-zero. It goes as follows. Let $f_N(A),\,A=(A_{\alpha\beta}$ is a generating functional 
(the density of the free energy) for correlation functions and define the order parameter
\be
Q_{\alpha\beta}^{(N)}=\frac{\partial f_N(A)}{\partial A_{\alpha\beta}}|_{A=0}=\frac{1}{N}\sum_{j=1}^N \langle \psi_j^{\alpha}\psi_j^{\beta}\rangle_N =0.
\ee 
So, if $N$ is finite, the order parameter $Q_{\alpha\beta}^{(N)}=0$. However, if we {\it first} compute the large $N$ limit,
$\lim_{N\to\infty}f_N(A)=f(A)$ and {\it then} compute the derivative,
\be
Q_{\alpha\beta}=\frac{\partial f(A)}{\partial A_{\alpha\beta}}|_{A=0},
\ee
we will find in this model that the order parameter $Q_{\alpha\beta}$ is nonvanishing and there is spontaneous symmetry breaking.

Various aspects of the non-diagonal solutions of the saddle point equations and Schwarzian corrections in the SYK  and related  models are discussed in \cite{9auth,SSS,AreVol},\cite{Georges00,Fu16,MScomments,Polchinski16,Garcia-garcia16,Gur-Ari18,Maldacena:2018lmt,AKTV,Belokurov:2018fnn,Wang:2018ijz,Kim:2019upg,Garcia-Garcia:2019poj}.

The model considered in this note is a 0-dimensional analogue of the conformal limit of the SYK model in which nondiagonal saddle points have been studied in \cite{AKTV,Wang:2018ijz} while the model investigated in \cite{AreVol} is an analogue of the total SYK model with replicas.

In the next section spontaneous symmetry breaking and the quasi-averages method are exposed. In Sect.3 the fermionic matrix model and the order parameter are discussed. In Sect.4 the cases $q=2$ and $q=4$ are considered  and spontaneous breaking of symmetry is shown. 

\section{Setup}
\subsection{Spontaneous symmetry breaking and quasi-averages}

A powerful method for investigation of the spontaneous symmetry breaking is the Bogoliubov method of quasi-averages, see \cite{Bog} and for a recent review \cite{Zagr}. The quasi-averages method proceeds as follows.
If, in a finite  volume $\Lambda$, one has a Hamiltonian $H$ which is invariant under a symmetry group, then one replaces
$H\to H+w B$ where $B$ is a suitable extensive operator breaking the symmetry and $w$ is a real parameter. Define the partition function $Z_{\Lambda}(w)=\Tr e^{-\beta(H+wB)}$,  $\beta>0$. Now we first take the infinite volume thermodynamic limit $\Lambda \to \mathbb{R}^d$ for the free energy 
\be
f(w)=-\lim_{\Lambda \to \mathbb{R}^d}\frac{1}{\beta|\Lambda |}\log Z_{\Lambda}(w)
\ee
and for correlation functions
and then compute the limit of the free energy  at $w=0$,\,\,$f(0)=\lim_{w\to +0}f(w)$ and the same for its derivatives $f^{'}(0),\,f^{''}(0),...$ and for the correlation functions
\be
\prec K \succ=\lim_{w\to +0}\lim_{\Lambda \to \mathbb{R}^d}\Tr (Ke^{-\beta(H+wB)})/Z_{\Lambda}(w)
\ee
where $K$ is some operator.
 It may happen that after taking the limit $w\to 0$
still there is a dependance  on $B$ in $f^{'}(0),\,f^{''}(0),...$ or in the correlation functions. In such a case one speaks of spontaneous symmetry breaking. In particular, $\prec K \succ$ is called the quasi-average of the operator $K$. The quantity $ \prec B\succ$ can be interpreted as an order parameter in the pure phase, selected by the symmetry breaking source.

\subsection{Example}
There is a simple example of application of quasi-averages for describing spontaneous symmetry breaking. Consider the partition function with the Higgs potential
\be
Z_N=\int_{\mathbb {R}}e^{-N(x^2-a^2)^2 }dx,\label{ZZN}
\ee
where $a>0,\,\,N=1,2,3,...$ There is a symmetry under transformation $x\to -x$, $<x>_N=0$ and moreover the free energy vanishes in the large $N$ limit:
\be
f_0=-\lim_{N\to\infty}\frac{1}{N}\log Z_N = 0.
\ee

Now we add a term breaking the symmetry:
\be
Z_N(w)=\int_{\mathbb {R}}e^{-NS}dx,\,\,\,\,\,\,S=(x^2-a^2)^2 +wx,\,\,w>0.
\ee
Then the free energy is
\be
f(w)=\lim_{N\to\infty}f_N(w)=-\lim_{N\to\infty}\frac{1}{N}\log Z_N(w)=(x^2_0-a^2)^2 +wx_0,
\ee
where $x_0=x_0(w)$ is a solution of the equation
\be
4x(x^2-a^2)+w=0\ee
with the lowest value of the action $S$.
Therefore we obtain 
\be
f'(w)|_{+0}=-a\ee
So there is a spontaneous symmetry breaking and quasi-average $\prec x \succ=-a$.
Note that in this consideration the limit of large $N$ plays a role of the thermodynamical limit.

Uniqueness of the solution $x_0=x_0(w)$, $w>0$, is used also when one performs the Legendre transform from the free energy $f(w)$ to the effective potential $\Gamma(\phi)=f(w)-w\phi$, $\phi=\partial _w f(w)$.

\section{Fermionic random matrix model}
\subsection{Model}
We consider  a polynomial function in anticommuting variables of the form \cite{AreVol}
\be
{\cal A}={\cal A}_0(J)+{\cal A}_1(A),
\ee
where
$$
{\cal A}_0(J)=\sum_{\alpha}\sum_{k_1<k_2<...<k_q}J_{k_1k_2...k_q}\chi_{k_1}^{\alpha}
\chi_{k_2}^{\alpha}...\chi_{k_q}^{\alpha},
$$
$$
{\cal A}_1(A)=\frac{i}2\sum_{\alpha,\beta,j}\chi_j^{\alpha}A_{\alpha\beta}
\chi_j^{\beta}.
$$
Here $\chi_j^{\alpha}$ are the Grassmann hermitean variables, $\{\chi_j^{\alpha},\chi_k^{\beta}\}=0,$\, $k,j=1,...N,\,\,\alpha,\beta=1,.. M$, $q$ is an even natural number,    and 
${\bf J}=(J_{k_1k_2...k_q})$ are  Gaussian random variables with
zero mean $<J_{k_1k_2...k_q}>=0$, and variance  is given by  $\langle J_{k_1k_2...k_q}^2\rangle =(q-1)!J^2/N^{q-1}$, $J>0$. The matrix $(A_{\alpha\beta})
$  is  antisymmetric  with complex entries. To get correlation functions we will differentiate the generating functional over   tensor type source $A_{\alpha\beta}
$ and then set $A_{\alpha\beta}=0.$

Although in this note we treat $A_{\alpha\beta}$ as auxiliary parameters in the generating functional, one can interpret them as coupling constants,
such models are discussed in 0-dim  case in \cite{AreVol} and in 1-dim in \cite{Maldacena:2018lmt,Garcia-Garcia:2019poj}.   
 
The generating  functional (partition function)  in the fermionic random matrix model is defined as  the Grassmann (Berezin) integral over anticommuting variables

\bea
Z_N(A)=\int d\mu ({\bf J})\,\int \, d\chi\,e^{i{\cal A}}.\label{Zchi}
\eea
Here    $d\mu ({\bf J})$ is a Gaussian probability measure. Mathematical questions of superanalysis are considered in \cite{VlaVol12}.
The integral over Grassmann variables  for $A_{\alpha\beta}=0$ vanishes unless $N$ is a multiple of 4,  so we assume that $N$ is a multiple of 4.
Note also that $Z_N(A)$ is a polynomial at $J^2$ and $A_{\alpha\beta}$ because there is only a finite number
of the Grassmann variables.

The free energy is defined as
\be
f_N (A)=-\frac{1}{N}\log Z_N(A).
\ee

We are interested in correlation functions
\bea
\label{2-poi}Q_{\alpha\beta}^{(N)}=\cfrac{\partial}{\partial A_{\alpha,\beta}}f_N(A)\big|_{A=0}=\frac1N\sum _{j=1}^N\,\langle\chi_j^{\alpha}
\chi_j^{\beta}\rangle_N.\eea

Note that $Q_{\alpha\beta}^{(N)}$ vanishes if $N$ is finite,
\be
Q_{\alpha\beta}^{(N)}=0,\,\,\,\alpha\neq\beta.\label{2-poi-m}
\ee
since the 2-point correlation function (\ref{2-poi}) factorizes into fermion 1-point functions, which vanish.
Another way to get the relation (\ref{2-poi-m}) is by notice that the function
${\cal A}_0(J)$ is invariant under the "gauge transformation" $\chi_j^{\alpha}\to\epsilon_{\alpha}\chi_j^{\alpha}$ where $\epsilon_{\alpha}=\pm 1$.
The relation (\ref{2-poi-m}) can be interpreted by saying that there is no symmetry breaking at finite $N$.

\subsection{Quasi-averages}

However if we first take the limit $N\to\infty$,
\be
f(A)=\lim_{N\to\infty}f_N(A)\label{fN}
\ee
and then compute the derivative, we could get a non-zero result for 2-point off-diagonal correlation function defined as 
\be\label{Quasi}
Q_{\alpha\beta}=\frac{\partial f(A)}{\partial A_{\alpha\beta}}\big|_{A=0}
\ee
This is the Bogoliubov quasi-averages approach to phase transitions and spontaneous symmetry breaking. 
The quantity $Q_{\alpha\beta}$ is called the quasi-average correlation function or the order parameter.

The expression \eqref{Zchi} can be written also by using the real variables
\bea
Z_N(A)=\left(\frac{N}{2\pi}\right)^{\frac{M(M-1)}{2}}\int d\Sigma\, dG\,e^{NS}\label{ZGS}
\eea
where
\bea
S=\log ( \mathrm{Pf}({\bf A}+i{\bf \Sigma}))+\sum_{\alpha<\beta}
\left[-\frac{J^2}{q}  G_{\alpha\beta}^q+i\Sigma_{\alpha\beta}G_{\alpha\beta}\right].
\label{Z0M}\eea
Here Pf is the Pfaffian, 
 ${\bf G}= (G_{\alpha\beta})$ and ${\bf \Sigma}= (\Sigma_{\alpha\beta})$ are antisymmetric matrices with real entries,  
 ${\bf A}= (A_{\alpha\beta})$ is antisymmetric complex matrix,  $q$ is an even integer. Note the presence of the factor $\exp\{-J^2  G_{\alpha\beta}^q/q\}$ which provides the convergence of the integral.
Note also that in \eqref{ZZN} and \eqref{ZGS} we use the different signs in front of the actions. We will show that for $q=2,\,4$ and $M=2$ the free energy $f({\bf A})$ in the limit $N\to\infty$ is equal to the action function $S_0$
computed at the saddle point $(G_{(0)},\,\Sigma_{(0)})$, i.e. $f({\bf A})=-S_0({\bf A})$. 
The saddle point equations read
\be\label{sp}
({\bf A}+i{\bf \Sigma})^{-1}=-{\bf G},\,\,\,\,\Sigma_{\alpha\beta}=-i\, J^2\,(G_{\alpha\beta})^{q-1}.
\ee

Since the gradient of the action function vanishes at the saddle point we obtain the following expression for the order parameter
\be
Q_{\alpha\beta}=-\frac{\partial S_0({\bf A})}{\partial A_{\alpha\beta}}\big|_{A=0}=i \big({\bf \Sigma}_{(0)}^{-1}\big)_{\alpha\beta}\big|_{A=0}
=G_{(0)\alpha\beta}|_{A=0}.
\ee

Let us note that the order parameter could depend on the way in which $A\to 0$ in complex domain.   
In this note we take $M=2$ and denote $A=A_{12}=a\e^{i\varphi}$, $a>0$. Then  we can define 
\be\label{Q12}
Q(\varphi)=\e^{-i\varphi}\frac{\partial}{\partial a}f(a\e^{i\varphi})\Big|_{a=0}.
\ee
If $Q(\varphi)$ is non-zero then one has  spontaneous symmetry breaking.

\section{Large $N$ asymptotics and quasi-averaging}
\subsection{q=2 model}
We start with $q=2,\,M=2$ case. 
\bea
Z_N(A)&=&\frac{N}{2\pi}\int_{\mathbb{R}} \, dy\,\left( A+iy\right)^N\,
\int_{\mathbb{R}} dx\exp\left\{N
\left[-\frac{J^2}{2}  x^2+ixy\right]\right\}\nn\\
&=&\frac{N}{2\pi}\int dx\, dy\,\,
\exp\left\{N
\left[-\frac{J^2}{2}  x^2+ixy+\ln \left( A+iy\right)\right]\right\}\label{Z0M}\eea
where we  use the parametrization
\be
\left(A_{\alpha\beta}\right)=
\left(
\begin{array}{cc}
0  &  A    \\
 -A &  0 
\end{array}
\right),\,\,\,\,\,\left(G_{\alpha\beta}\right)=
\left(
\begin{array}{cc}
0  &  x    \\
 -x &  0 
\end{array}
\right)
,\,\,\,\,\,\left(\Sigma_{\alpha\beta}\right)=
\left(
\begin{array}{cc}
0  &  y    \\
 -y &  0 
\end{array}
\right)
\ee
Here $A$ is a complex parameter with small enough $|A|$.  
The stationary point equations are
\bea
-J^2 x+i y&=&0\\
\frac{i}{A+i y}+i x&=&0\eea
The solutions to these equations have the form
  
   \bea
x_1&=&\frac{-A-i \sqrt{4 J^2-A^2}}{2 J^2},\\
y_{1} & = & \frac{i}{2} \left(A+i \sqrt{4
   J^2-A^2}\right),\\   x_2& = & \frac{-A+i \sqrt{4 J^2-A^2}}{2 J^2},\\\,\,
   y_2& = & -\frac{i}{2}  \left(-A+i \sqrt{4
   J^2-A^2}\right)\eea
   The action at the stationary points reads 
   \bea
S_{1}(A)&=&\log
   \left(\frac{1}{2} \left(A-i \sqrt{4
   J^2-A^2}\right)\right)+\frac{1}{4} \left(\frac{A^2}{J^2}-2+\frac{i
   A \sqrt{4 J^2-A^2}}{J^2}\right)\\
   S_{2}(A)&=&\log \left(\frac{1}{2} \left(A+i \sqrt{4
   J^2-A^2}\right)\right)+\frac{1}{4} \left(\frac{A^2}{J^2}-2-\frac{i
   A \sqrt{4 J^2-A^2}}{J^2}\right)\eea
   We consider these action for $A=a\,\e^{i\varphi}, \,\,\,a>0$
   and get
   \bea
  S_{1}(a,\varphi)&=&\log
   \left(\frac{1}{2} \left(a\,\e^{i\varphi}-i \sqrt{4
   J^2-a^2\,\e^{2i\varphi}}\right)\right)+\frac{1}{4} \left(\frac{a^2\,\e^{2i\varphi}}{J^2}-2+\frac{i
   a\,\e^{i\varphi} \sqrt{4 J^2-a^2\,\e^{2i\varphi}}}{J^2}\right)\nn\\\label{q2A1}
    \\
   S_{2}(a,\varphi)&=&\log
   \left(\frac{1}{2} \left(a\,\e^{i\varphi}-i \sqrt{4
   J^2-a^2\,\e^{2i\varphi}}\right)\right)+\frac{1}{4} \left(\frac{a^2\,\e^{2i\varphi}}{J^2}-2-\frac{i
   a\,\e^{i\varphi} \sqrt{4 J^2-a^2\,\e^{2i\varphi}}}{J^2}\right)\nn\\ \label{q2A2}
   \eea
   
\begin{figure}[h!]
\centering
  \includegraphics[width=6cm]{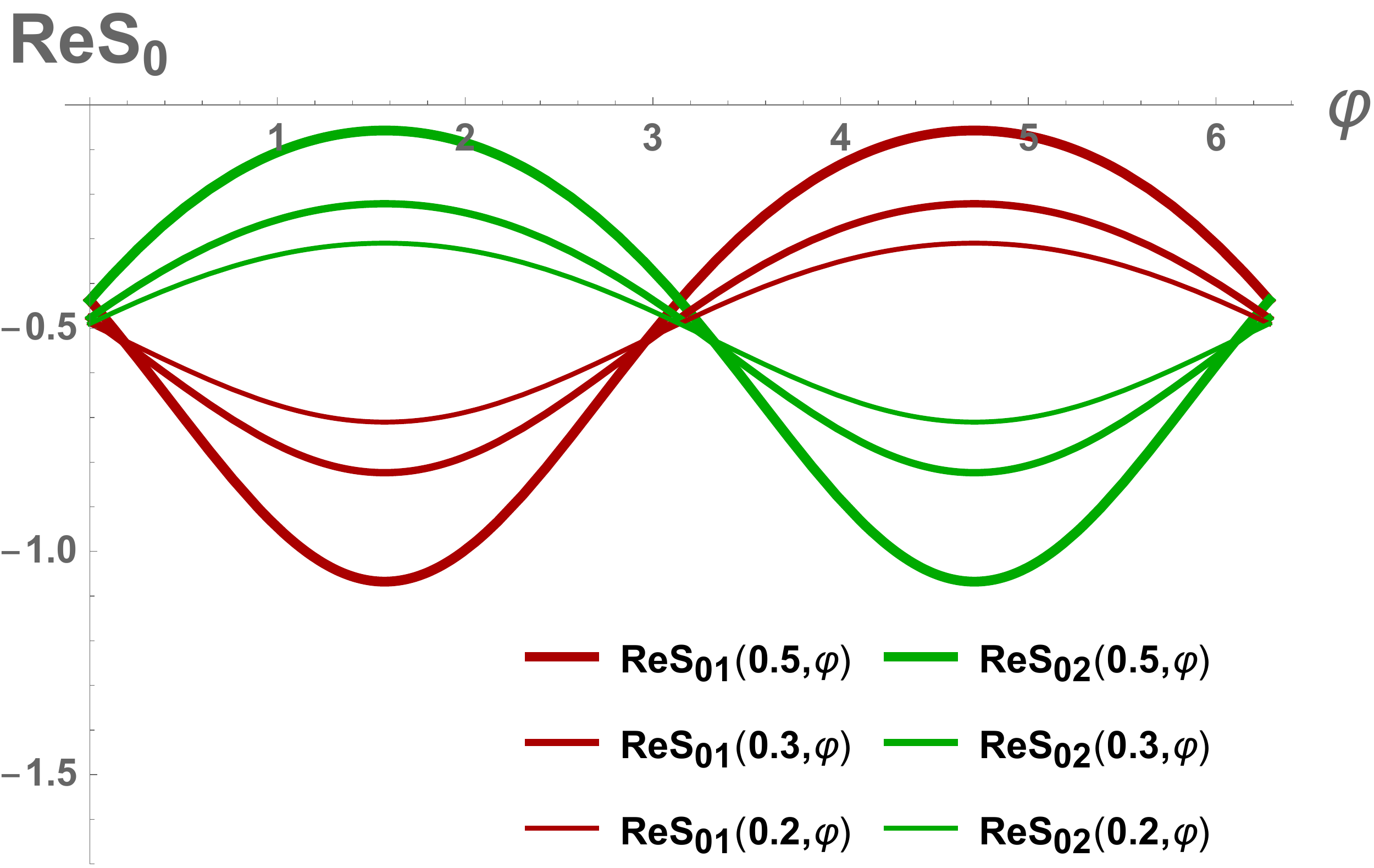}\,\,\,\,\,\,\,\,\,\,\,\,\,\,
   \includegraphics[width=6cm]{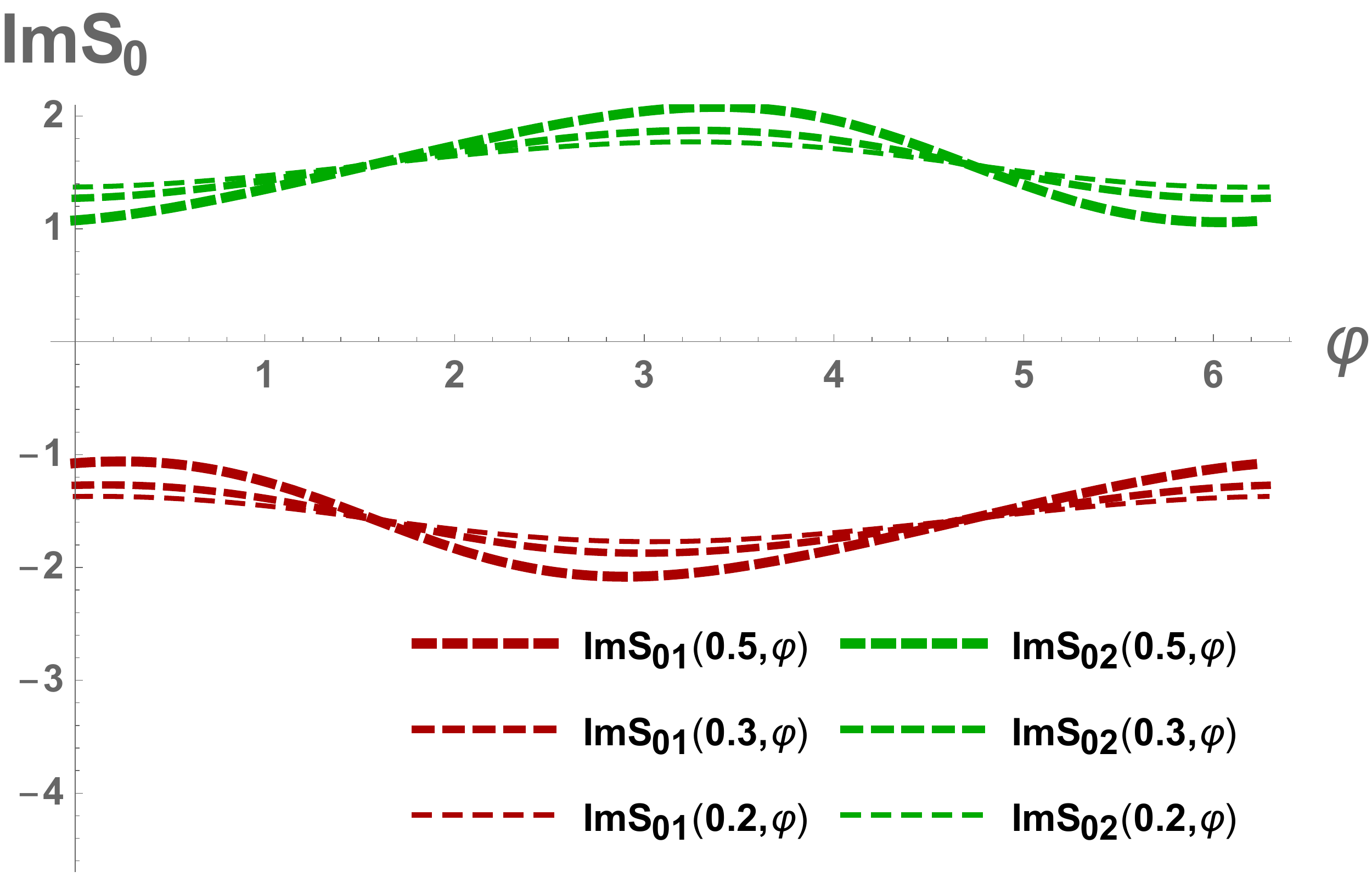}\\
   {\bf A}\,\,\,\,\,\,\,\,\,\,\,\,\,\,\,\,\,\,\,\,\,\,\,\,\,\,\,\,\,\,\,\,\,\,\,\,\,\,\,\,\,\,\,\,\,\,\,\,\,\,\,\,\,\,\,\,{\bf B}

  \caption{ Plots of real  ({\bf A}) and imaginary ({\bf B}) parts of the actions on the stationary point,  $S_{0,i}(a,\varphi)$ for $J=1$, different $a<2$ and $0<\varphi<2\pi$.  $i=1,2$ corresponds to the first and second stationary points and
          are presented by red and green color, respectively. 
          }
  \label{Fig:action-q2}
\end{figure}

   In Fig.\ref{Fig:action-q2} we plot the real and imaginary parts of  actions \eqref{q2A1} and \eqref{q2A2}
   for $0<a<2$.  We see that the real part of the first solution for $0<\varphi<\pi$ is less then the real part of the  action on the second one 
          and the second solution dominates.
          For $\pi<\varphi<2\pi $  the first solution dominates. 
          
\begin{figure}[h!]
 \centering
        \includegraphics[width=10cm]{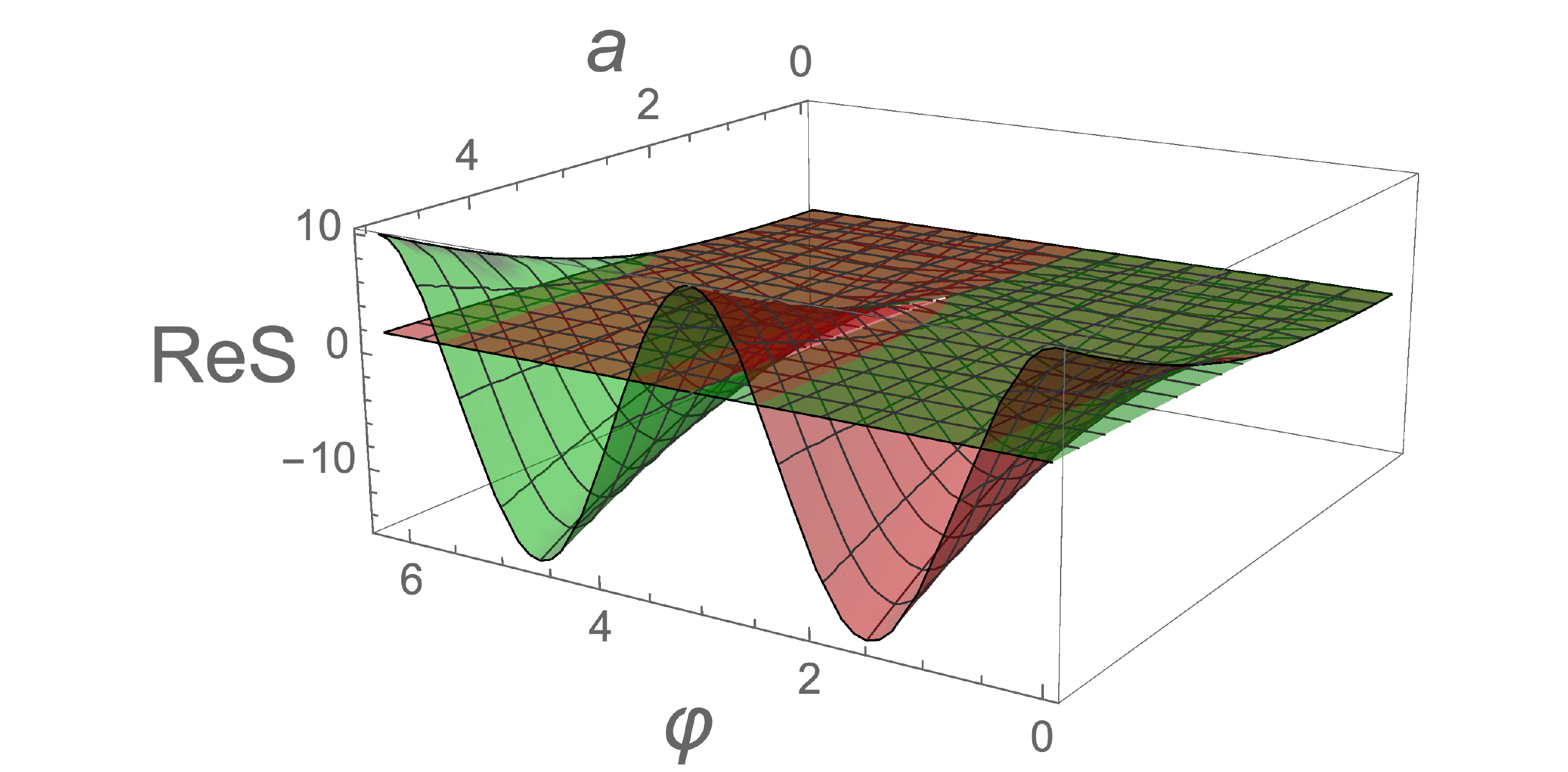}
   \includegraphics[width=2cm]{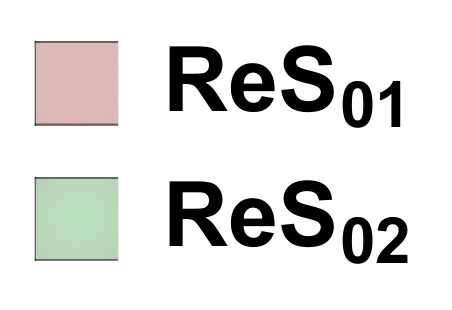}\,\,\,\,\,\,\,\\
\caption{ Plots for real parts of the actions at the stationary point,  $ReS_{0,i}(a,\varphi)$ for $J=1$,  $0<a<4$ and $0<\varphi<2\pi$. $i=1,2$ corresponds to the first and second stationary points and
          are presented by red and green color respectively.
          }
  \label{Fig:action-q2-3D}
\end{figure}

          In  Fig.\ref{Fig:action-q2-3D} we show  the real  parts of the actions also for $a>2$. 
          We see the  order of dominance is changed for $a>2$.

 Therefore the free energy for  $A=ae^{i\varphi}$ ($0<a<2$) is given by 
\be
f(a,\varphi)=
\left\{
\begin{array}{ccc}
 -S_{2}(a,\varphi) & \mbox{for}  & 0<\varphi <\pi \\
-S_{1}(a,\varphi) & \mbox{for}   & \pi <\varphi <2\pi\\
\end{array}
\right.\,\,\,\,\,\,\,\,
\ee

Now we can evaluate the value of the order parameter $Q$.
We have

\bea
 \frac{\partial}{\partial a}S_{1}(a,\varphi)&=&\frac{2}{a-i e^{-i \varphi } \sqrt{4
   J^2-a^2 e^{2 i \varphi }}}\label{HS1}\\
  \frac{\partial}{\partial a}S_{2}(a,\varphi)&=& \frac{2}{a+i e^{-i \varphi } \sqrt{4
   J^2-a^2 e^{2 i \varphi }}}\label{HS2}\eea

Now we can obtain the order parameter  $Q$. We finally get
\bea
Q(\varphi)=-\left\{
\begin{array}{cc}
e^{-i \varphi } \frac{\partial}{\partial a}S_{2}(a,\varphi) \Big|_{a=0} & 
   =  \frac{i}{J}, \,\,\,\,\, 0<\varphi<\pi \\\,&\\
 e^{-i \varphi } \frac{\partial}{\partial a}S_{1}(a,\varphi)  \Big|_{a=0} &
   = -\frac{i}{J} , \,\,\,\,\,\,   \pi<\varphi<2\pi \,.
\end{array}
\right.\label{Qphi2}
\eea

The last formula shows that there is spontaneous symmetry breaking and the off-diagonal correlation function is nonvanishing.

Note that the partition function (\ref{Z0M}) can be written using the Hermite polynomials \cite{AreVol}
\be
Z_N(A)=2\left(\frac{J}{\sqrt {2N}}\right)^N\,H_N\left(\frac{A}{J}\,\sqrt {\frac{N}{2}}\right).
\ee
The asymptotic behaviour of the Hermite and more general polynomials in the complex domain is evaluated in \cite{Deift}.

\subsection{q=4 model}

Our starting point is 
\be
 Z_N(A)=\frac{N}{2\pi}\int dy\int dx\,(iy +A)^N
\exp\{N (-J^2\frac{x^4}{4}+ixy)\}\ee
$$=\frac{N}{2\pi}J^{N/2}\int dy\int dx\,
\exp\{NS\}
,
$$
where 
\be
S=-\frac{x^4}{4}+ixy+
\log (iy+\tilde{A})\label{act4}
\ee
and
$\tilde{ A}=A/\sqrt{J}.$

Here $x,y$ are real variables and  $A$ is a complex parameter, $J>0$ and $N$ is a natural number.

\begin{figure}[h!]
 \centering
  
    \includegraphics[width=6cm]{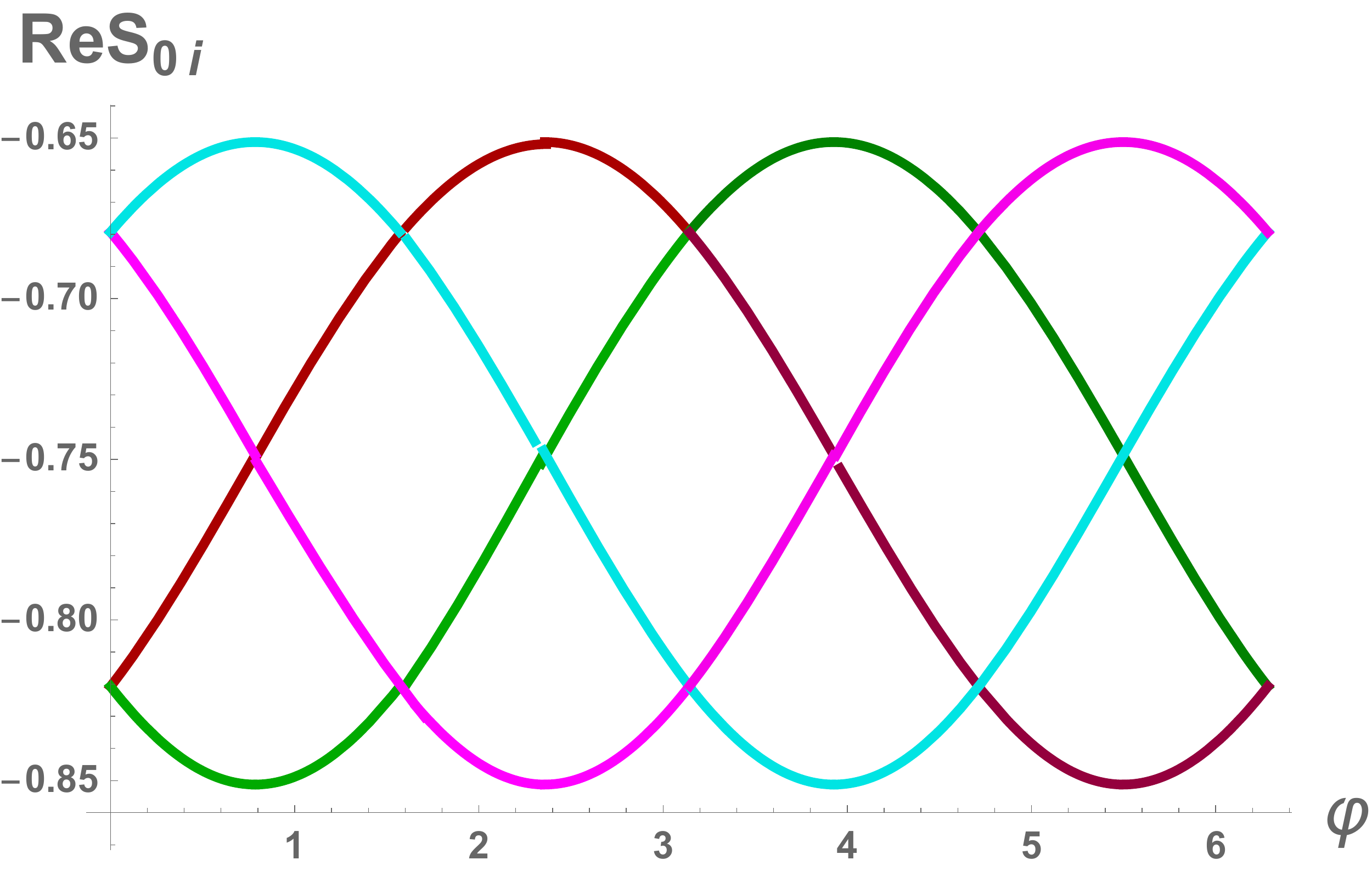}   \,\,\,\,\,\,\,\includegraphics[width=2cm]{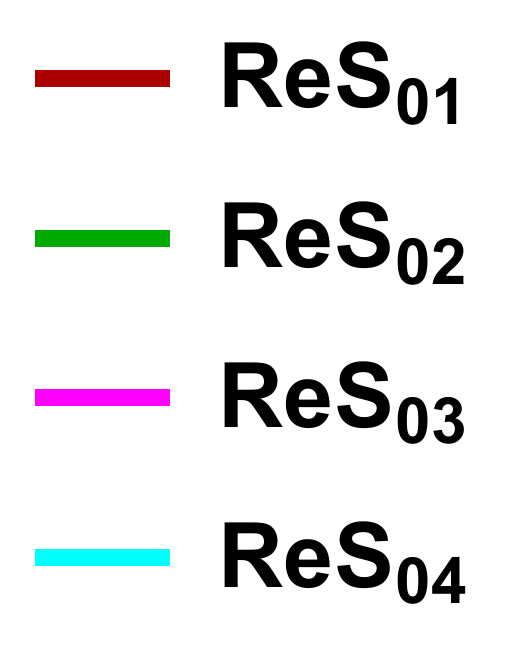}
   \caption{{\bf A}.  Plots for real parts of the actions on the stationary point functions,  $ReS_{0,i}(a,\varphi)$ for $J=1$,  $a=0.1$ and $0<\varphi<2\pi$, $i=1,2,3,4$ corresponding to the first, second, third and fourth  stationary points
          are presented by darker red, green, magenta and cyan  color, respectively.
     }
  \label{Fig:action-q4}
\end{figure}

The saddle points for $N\to\infty$ are derived from equations
\bea
- x^3+i y&=&0\label{root4}\\
\frac{1}{ \tilde{A}+i y}+ x&=&0,\eea
There are 4 roots and we numerate them by index $i$.
The values of the action \eqref{act4} at these roots are denoted by 
$S_{0i}, i=1,2,3,4$. The real parts of the action on these roots are degenerate in the sense that they 
are pairwise equal for real $A$. To remove this degeneracy as in the case $q=2$ we consider the complex $A=a e^{i \varphi},\,\,a>0$.
The values of the real parts of the actions are presented in Fig.\ref{Fig:action-q4} for a particular $a=0.1$ and $J=1$.
     We see that the real part of the fourth solution for $0<\varphi<\pi/2$ is bigger than the real parts of the  action on other solutions.
          For $\pi/2<\varphi<\pi $  the first solution dominates,   for $\pi<\varphi<3/2\pi $  the second solution dominates and  for $3/2\pi<\varphi<2\pi $   the  third one.

\begin{figure}[h!]
  \centering\includegraphics[width=6.5cm]{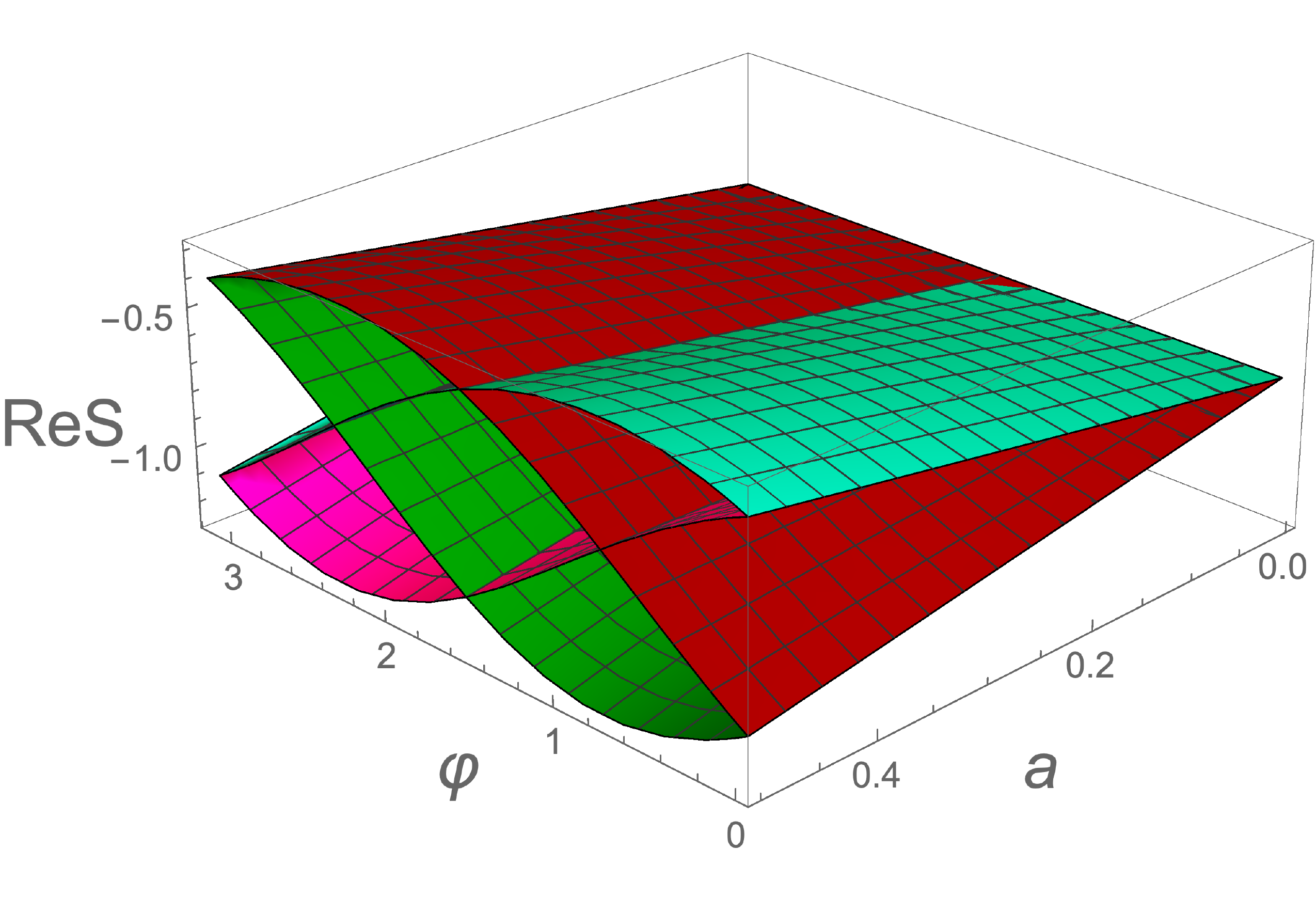}
   \includegraphics[width=1cm]{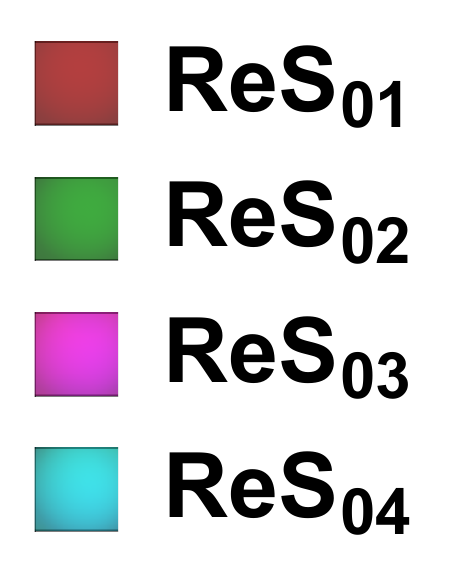}
\centering\includegraphics[width=6cm]{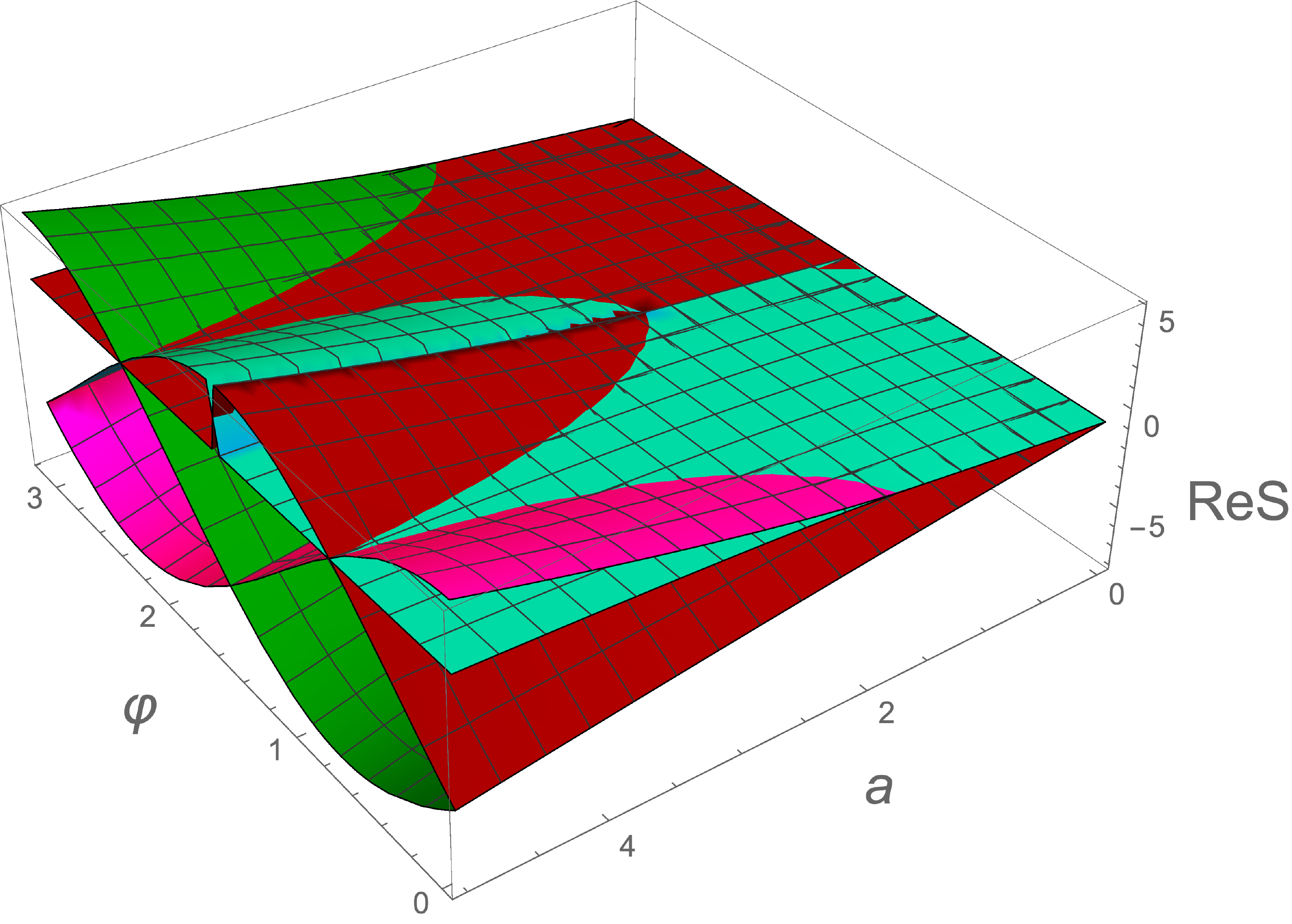}  
{\bf A}\,\,\,\,\,\,\,\,\,\,\,\,\,\,\,\,\,\,\,\,\,\,\,\,\,\,\,\,\,\,\,\,\,\,\,\,\,\,\,\,\,\,\,\,\,\,\,\,\,\,{\bf B}

  \caption{{\bf A}) 3D plots for functions $Re S_{0,1}( a\,e^{i\varphi})$ (red), $Re S_{0,2}( a\,e^{i\varphi})$ (green),
  $Re S_{0,3}( a\,e^{i\varphi})$ (magenta)  and $Re S_{0,4}( a\,e^{i\varphi})$ (cyan) at $J=1$, small $a$ and $0<\varphi<\pi$ show that the complex $A=a\,e^{i\varphi}$ removes the degeneracy of the real part of the action. {\bf B}) The same plot as on A), but for large variety of $a$, $0<a<5$.   
   }
  \label{Fig:3Dact-q4}
\end{figure}

The absence of degeneracy  for complex $A$ is also shown in the 3D plot  in Fig.\ref{Fig:3Dact-q4}. The plot in Fig.\ref{Fig:3Dact-q4}.B
shows that the order of dominance is changed at $a=a_0>1$.

Calculating  the first derivatives of the actions $S_{0,i}(a,\varphi)$
we get 
\bea \cfrac{\partial}{\partial a}S_{0,1}(a,\varphi)\Big|_{a=0}&=&-\frac{(1-i) e^{i \varphi }}{\sqrt{2}}\nn\\
\cfrac{\partial}{\partial a}S_{0,2}(a,\varphi)\Big|_{a=0}&=&-\frac{(1+i) e^{i \varphi }}{\sqrt{2}}\nn\\
\cfrac{\partial}{\partial a}S_{0,3}(a,\varphi)\Big|_{a=0}&=&\frac{(1+i) e^{i \varphi }}{\sqrt{2}}\nn\\
\cfrac{\partial}{\partial a}S_{0,4}(a,\varphi)\Big|_{a=0}&=&\frac{(1-i) e^{i \varphi }}{\sqrt{2}}\nn
\eea
To compute the value of the order parameter we have to take the  derivative of  the action
at the branch with the maximal real part in the considered domain of $\varphi$.  Finally we obtain
\be
Q(\varphi)=
\left\{
\begin{array}{ccc}
 -\frac{(1-i) }{\sqrt{2}}& \,\,\,\,\,\,\,\mbox{for} & 0<\varphi <\pi/2 \\
\frac{(1-i) }{\sqrt{2}} & \,\,\,\,\,\,\,\mbox{for}   & \pi/2 < \varphi <\pi\\
 \frac{(1+i)}{\sqrt{2}}& \,\,\,\,\,\,\,\mbox{for}   & \pi <\varphi <3/2\pi\\
 -\frac{(1+i) }{\sqrt{2}} & \,\,\,\,\,\,\,\mbox{for}  & 3\pi/2 <\varphi <2\pi.
\end{array}
\right.\ee

Nonvanishing of the order parameter means that there is spontaneous symmetry breaking.

\section{Conclusion}

We have considered the fermionic random matrix model     which is a  0-dimensional version of the  SYK model with a finite  number of replicas. It is shown  that although the 2-point non-diagonal correlation function in the fermionic matrix model vanishes at finite $N$, it does not vanish in the large $N$ limit due to spontaneous symmetry breaking. It would be interesting to study whether this  phenomena may occur in  other models.

\section*{Acknowledgements}  We thank  M.~Khramtsov and  V.~Zagrebnov  for useful discussions.


\begin{thebibliography}{99}
\bibitem{Landau} L.~D.~Landau, {\it On the theory of phase transitions},  Zh. Eksp. Teor. Fiz. 7 (1937) 19-32.\\ Landau L.D. In "Collected Papers" (Nauka, Moscow, 1969),
Vol. 1, pp. 234-252.
 \bibitem{Bog} 
 N. ~N. ~Bogoliubov, {\it Lectures on quantum statistics, volume 2: Quasi-Averages}.
Gordon and Breach Sci. Publ., 1970.\\ N. N. Bogoliubov, Collection of Scientific Works in Twelve Volumes: Statistical Mechanics, vol.6. Nauka, Moscow, 2007.
\bibitem{Numbu}  Y.~ Nambu, {\it Quasi-Particles and Gauge Invariance in the Theory of Superconductivity},
Phys. Rev. 117 (1960) 648,  
\bibitem{Weinb}  S.~ Weinberg, {\it The quantum theory of fields. Vol. 2: Modern applications},
Cambridge University Press, 1996
\bibitem{MezPar} 
 M.~Mezard, G.~Parisi and M.~Virasoro, {\it Spin Glass Theory and beyond,}
 World Scientific, 1987 
 



  \bibitem{9auth}  J.~S.~Cotler {\it et al.},
  {\it Black Holes and Random Matrices,}
  JHEP {\bf 1705}, 118 (2017),
  [arXiv:1611.04650 [hep-th]].
     \bibitem{AreVol}   I.~Ya.~Aref'eva and I.~V.~Volovich,
  {\it Notes on the SYK model in real time,}
  Theor.\ Math.\ Phys.\  197 (2018) 1650, 
 [arXiv:1801.08118 [hep-th]]

\bibitem{SSS}  P. Saad, S. H. Shenker and D. Stanford, {\it A semiclassical ramp in SYK and in gravity},
arXiv:1806.06840 [hep-th].



\bibitem{SY} S. Sachdev and J. Ye, "Gapless spin fluid ground state in a random, quantum
Heisenberg magnet," Phys. Rev. Lett. 70 (1993) 3339, [arXiv:cond-mat/9212030
[cond-mat]].
\bibitem{AK}
A.~Kitaev, {\it A simple model of quantum holography,}  KITP strings
  seminar and Entanglement 2015 program, (Feb. 12, April 7, and May 27, 2015) .
  http://online.kitp.ucsb.edu/online/entangled15/.
\bibitem{Sachdev:2015}
  S.~Sachdev,
  {\it Bekenstein-Hawking Entropy and Strange Metals,}
  Phys.\ Rev.\ X {\bf 5}, no. 4, 041025 (2015),
  [arXiv:1506.05111 [hep-th]].
  
    \bibitem{Zagr} Walter F.~ Wreszinski and Valentin A. ~Zagrebnov, {\it Bogoliubov quasi-averages: spontaneous symmetry breaking and algebra of fluctuations},  Theor.\ Math.\ Phys.\ 194  (2018) 157-188,  [arXiv:1704.00190[math-ph]]
  \bibitem{Georges00}
A.~Georges, O.~Parcollet and S.~Sachdev, {\it Quantum fluctuations of a
  nearly critical Heisenberg spin glass,}
  Phys.\ Rev.\ B  {\bfseries 63} (Apr., 2001) 134406
  [arXiv:  cond-mat/0009388].

 \bibitem{Polchinski16} 
  J.~Polchinski and V.~Rosenhaus,
  {\it The Spectrum in the Sachdev-Ye-Kitaev Model,}
  JHEP {\bf 1604}, 001 (2016),
  [arXiv:1601.06768 [hep-th]].
 

\bibitem{Fu16} 
  W.~Fu and S.~Sachdev,
  {\it Numerical study of fermion and boson models with infinite-range random interactions,}
  Phys.\ Rev.\ B {\bf 94}, no. 3, 035135 (2016)
  [arXiv:1603.05246 [cond-mat.str-el]].


\bibitem{MScomments} 
  J.~Maldacena and D.~Stanford,
  {\it Remarks on the Sachdev-Ye-Kitaev model,}
  Phys.\ Rev.\ D {\bf 94}, no. 10, 106002 (2016)
  [arXiv:1604.07818 [hep-th]].
 

 \bibitem{Garcia-garcia16} 
  A.~M.~Garcia-Garcia and J.~J.~M.~Verbaarschot,
  {\it Spectral and thermodynamic properties of the Sachdev-Ye-Kitaev model,}
  Phys.\ Rev.\ D {\bf 94}, no. 12, 126010 (2016)
  [arXiv:1610.03816 [hep-th]].
 
  \bibitem{Gur-Ari18} 
  G.~Gur-Ari, R.~Mahajan and A.~Vaezi,
  {\it Does the SYK model have a spin glass phase?,}
  arXiv:1806.10145 [hep-th].

   
    \bibitem{AKTV} 
  I.~Aref'eva, M.~Khramtsov, M.~Tikhanovskaya and I.~Volovich,
  {\it Replica-nondiagonal solutions in the SYK model,}
  arXiv:1811.04831 [hep-th].
  \bibitem{Belokurov:2018fnn} 
  V.~V.~Belokurov and E.~T.~Shavgulidze,
  {\it Simple rules of functional integration in the Schwarzian theory: SYK correlators,}
  arXiv:1811.11863 [hep-th].
  \bibitem{Wang:2018ijz} 
  H.~Wang, D.~Bagrets, A.~L.~Chudnovskiy and A.~Kamenev,
  {\it On the replica structure of Sachdev-Ye-Kitaev model,}
  arXiv:1812.02666 [hep-th].
  \bibitem{Kim:2019upg} 
  J.~Kim, I.~R.~Klebanov, G.~Tarnopolsky and W.~Zhao,
  {\it Symmetry Breaking in Coupled SYK or Tensor Models,}
  arXiv:1902.02287 [hep-th].
 
  
  \bibitem{Maldacena:2018lmt}
  J.~Maldacena and X.~L.~Qi,
  {\it Eternal traversable wormhole,}
  arXiv:1804.00491 [hep-th].
  \bibitem{Garcia-Garcia:2019poj} 
  A.~M.~Garcia-Garcia, T.~Nosaka, D.~Rosa and J.~J.~M.~Verbaarschot,
  {\it Quantum chaos transition in a two-site SYK model dual to an eternal traversable wormhole,}
  arXiv:1901.06031 [hep-th].


\bibitem{VlaVol12} V.S.Vladimirov, I.V.Volovich, {\it Differential calculus},  Theor. Math. Phys., 59 (1984), 317-335;
 "Superanalysis. II. Integral calculus", Theor. Math. Phys., 60 (1984), 743-765. 
 
\bibitem{Deift} P.~ Deift, T.~ Kriecherbauer
K. ~T-R. ~Mclaughlin, 
S. ~Venakides and 
X.~ Zhou, {\it  Strong Asymptotics of Orthogonal Polynomials
with Respect to Exponential Weights},
Comm. on Pure and Applied Math., Vol. LII,  (1999) 1491-1552.
  \end{thebibliography}
\end{document}